\documentclass[a4paper,useAMS,usenatbib,fleqn]{mn2e}

\usepackage{color}
\usepackage{natbib}
\usepackage{amsfonts}
\usepackage{amsmath}
\usepackage{aas_macros}
\usepackage{threeparttable}
\usepackage{url}
\usepackage[printwatermark]{xwatermark}
\usepackage{xcolor}
\usepackage{graphicx}

\title[Strong-lensing of GW170814?]{Deep and rapid observations of strong-lensing galaxy clusters within the sky localization of GW170814}

\author[G.\ P.\ Smith et al.]
       {G.\ P.\ Smith,$\!^1$\thanks{E-mail: gps@star.sr.bham.ac.uk}
        M.\ Bianconi,$\!^1$
        M.\ Jauzac,$\!^{2,3,4}$
        J.\ Richard,$\!^5$
        A.\ Robertson,$\!^3$\newauthor
        C. P. L. Berry,$\!^6$
        R.\ Massey,$\!^2$
        K.\ Sharon,$\!^7$
        W.\ M.\ Farr,$\!^{8,9}$
        J.\ Veitch$^{10}$
        \\\\
$^1$ School of Physics and Astronomy, University of Birmingham, Birmingham, B15 2TT, England\\
$^2$ Centre for Extragalactic Astronomy, Department of Physics, Durham University, Durham DH1 3LE, England\\
$^3$ Institute for Computational Cosmology, Durham University, South Road, Durham DH1 3LE, England\\
$^4$ Astrophysics and Cosmology Research Unit, School of Mathematical Sciences, University of KwaZulu-Natal, Durban 4041, South Africa\\
$^5$ Univ Lyon, Univ Lyon1, Ens de Lyon, CNRS, Centre de Recherche Astrophysique de Lyon UMR5574, F-69230, Saint-Genis-Laval, France\\
$^6$ Center for Interdisciplinary Exploration \& Research in Astrophysics (CIERA), Northwestern University, Evanston, IL 60208, USA\\
$^7$ Department of Astronomy, University of Michigan, 1085 S.\ University Avenue, Ann Arbor, MI 48109, USA\\
$^8$ Department of Physics and Astronomy, Stony Brook University, Stony Brook, NY 11794, USA\\
$^9$ Center for Computational Astrophysics, Flatiron Institute, 162 Fifth Avenue, New York, NY 10010, USA\\
$^{10}$ School of Physics and Astronomy, University of Glasgow, Glasgow, G12 8QQ, Scotland\\
}

\begin{document}

\date{Accepted 2019 February 13. Received 2019 February 13; in original form 2018 May 18.}
\pagerange{\pageref{firstpage}--\pageref{lastpage}} \pubyear{2019}
\maketitle
\label{firstpage}

\newcommand{\red}{\textcolor{red}}
\newcommand{\blue}{\textcolor{blue}}
\newcommand{\norv}[1]{{\textcolor{blue}{#1}}}
\newcommand{\norvnew}[1]{{\textcolor{red}{#1}}}
\def\hkpc{\ensuremath{h^{-1}~\mathrm{kpc}}}
\def\hMpc{\ensuremath{h^{-1}~\mathrm{Mpc}}}
\def\Mpc{\ensuremath\mathrm{Mpc}}
\def\Mvir{\ensuremath{M_\mathrm{vir}}}
\def\cvir{\ensuremath{c_\mathrm{vir}}}
\def\rvir{\ensuremath{r_\mathrm{vir}}}
\def\Dvir{\ensuremath{\Delta_\mathrm{vir}}}
\def\rsc{\ensuremath{r_\mathrm{sc}}}
\def\rhoc{\ensuremath{\rho_\mathrm{crit}}}
\def\Msol{\ensuremath\mathrm{M_\odot}}
\def\hMsol{\ensuremath{h^{-1}\mathrm{M_\odot}}}
\def\h70Msol{\ensuremath{h_{70}^{-1}\mathrm{M_\odot}}}
\def\ergs{\ensuremath\mathrm{erg\,s^{-1}}}
\def\Mgas{\ensuremath{M_\mathrm{gas}}}
\def\Mhse{\ensuremath{M_\mathrm{HSE}}}
\def\Mp{\ensuremath{M_\mathrm{Planck}}}
\def\Mwl{\ensuremath{M_\mathrm{WL}}}
\def\Mfh{\ensuremath{M_{500}}}
\def\rfh{\ensuremath{r_{500}}}
\def\Tx{\ensuremath{T_X}}
\def\Om{\ensuremath{\Omega_\mathrm{M}}}
\def\Ol{\ensuremath{\Omega_\Lambda}}
\def\keV{\ensuremath\mathrm{keV}}
\def\kpc{\ensuremath\mathrm{kpc}}
\def\kms{\ensuremath\mathrm{km\,s^{-1}}}
\def\ls{\ensuremath{\hbox{\rlap{\hbox{\lower4pt\hbox{$\sim$}}}\hbox{$<$}}}}
\def\gs{\ensuremath{\hbox{\rlap{\hbox{\lower4pt\hbox{$\sim$}}}\hbox{$>$}}}}
\def\ds{\ensuremath{D_\mathrm{S}}}
\def\dls{\ensuremath{D_\mathrm{LS}}}
\def\dsi{\ensuremath{D_{\mathrm{S},i}}}
\def\dlsi{\ensuremath{D_{\mathrm{LS},i}}}
\def\dsj{\ensuremath{D_{\mathrm{S},j}}}
\def\dlsj{\ensuremath{D_{\mathrm{LS},j}}}
\def\dsi{\ensuremath{D_{\mathrm{S},i}}}
\def\zs{\ensuremath{z_{S}}}
\def\ks{\ensuremath\mathrm{ks}}
\def\betaP{\ensuremath{\beta_\mathrm{P}}}
\def\betaX{\ensuremath{\beta_\mathrm{X}}}
\def\zs{\ensuremath{z_\mathrm{S}\,}}
\newcommand{\perGpcyr}{\ensuremath{\mathrm{Gpc}^{-3}\,\mathrm{yr}^{-1}}}
\newcommand{\be}{\begin{equation}}
\newcommand{\ee}{\end{equation}}
\newcommand{\ba}{\begin{eqnarray}}
\newcommand{\ea}{\end{eqnarray}}

\newcommand{\cplb}[1]{{\textcolor{cyan}{\sf{[CPLB: #1]}} }}
\newcommand{\jv}[1]{{\textcolor{green}{\sf{[JV: #1]}} }}
\newcommand{\gps}[1]{{\textcolor{purple}{\sf{[GPS: #1]}} }}
\newcommand{\ar}[1]{{\textcolor{orange}{\sf{[AR: #1]}} }}

\newcommand{\edit}[1]{\textbf{#1}}

\addtolength\topmargin{-12mm}

\begin{abstract}
We present observations of two strong-lensing galaxy clusters located
within the $90$ per cent credible sky localization maps released
following LIGO-Virgo's discovery of the binary black hole (BH-BH)
gravitational wave (GW) source GW170814.  Our objectives were (1) to
search for candidate electromagnetic (EM) counterparts to GW170814
under the hypothesis that it was strongly-lensed, and thus more
distant and less massive than inferred by LIGO-Virgo, and (2) to
demonstrate the feasibility of rapid target of opportunity
observations to search for faint lensed transient point sources in
crowded cluster cores located within GW sky localizations.  Commencing
$20$ hours after discovery, and continuing over $12$ nights, we
observed Abell~3084 ($z=0.22$) and SMACS\,J0304.3$-$4401 ($z=0.46$)
with GMOS on the Gemini-South telescope, and Abell~3084 with MUSE on
ESO's Very Large Telescope.  We detect no candidate EM counterparts in
these data.  Calibration of our photometric analysis methods using
simulations yield $5\sigma$ detection limits for transients in
difference images of the cores of these clusters of $i=25$.  This is
the most sensitive photometric search to date for counterparts to GW
sources, and rules out the possibility that GW170814 was lensed by
these clusters with a kilonova-like EM counterpart.  Based on the
detector frame masses of the compact objects, and assuming that at
least one Neutron Star (NS) is required in the merging system to
produce a kilonova-like counterpart, implies that GW170814 was neither
a NS-NS nor NS-BH merger at $z>8$ lensed by either of these clusters.
Also, in the first ever emission line search for counterparts to GW
sources, we detected no lines down to a $5\sigma$ detection limit of
$5\times10^{-17}{\rm erg\,s^{-1}\,cm^{-2}}$.
\end{abstract}


\makeatletter

\def\doi{\begingroup
  % The following isn't just \dospecials, because that includes \ , \{, and \}
  \let\do\@makeother \do\\\do\$\do\&\do\#\do\^\do\_\do\%\do\~
  \@ifnextchar[%]
    {\@doi}
    {\@doi[]}}
\def\@doi[#1]#2{%
  \def\@tempa{#1}%
  \ifx\@tempa\@empty
    \href{http://dx.doi.org/#2}{doi:#2}%
  \else
    \href{http://dx.doi.org/#2}{#1}%
  \fi
  \endgroup
}

%
\def\eprint#1#2{%
  \@eprint#1:#2::\@nil}
\def\@eprint@arXiv#1{\href{http://arxiv.org/abs/#1}{{\tt arXiv:#1}}}
\def\@eprint@dblp#1{\href{http://dblp.uni-trier.de/rec/bibtex/#1.xml}{dblp:#1}}
\def\@eprint#1:#2:#3:#4\@nil{%
  \def\@tempa{#1}%
  \def\@tempb{#2}%
  \def\@tempc{#3}%
  \ifx\@tempc\@empty
    \let\@tempc\@tempb
    \let\@tempb\@tempa
  \fi
  \ifx\@tempb\@empty
    \def\@tempb{arXiv}%
  \fi
  \@ifundefined{@eprint@\@tempb}
    {\@tempb:\@tempc}
    {
      \expandafter\expandafter\csname @eprint@\@tempb\endcsname\expandafter{\@tempc}}%
}

%
\def\mniiiauthor#1#2#3{%
  \@ifundefined{mniiiauth@#1}
    {\global\expandafter\let\csname mniiiauth@#1\endcsname\null #2}
    {#3}}

\makeatother


\section{Introduction}\label{sec:intro}

Observational astronomy gained a new tool with the first direct
detection of GWs \citep{GW150914}. GWs have already provided new
insights in to the properties of compact binaries and the nature of
gravity \citep[e.g.,][]{Interpretation150914, Abbott2016-GW150914-tgr,
  GW170104detect, O2-Pops} that complement those accessible to EM
observations.  In the case of GW170817 \citep{GW170817detect}, the
first GW signal from a binary neutron star (NS-NS) coalescence was
followed by observations of a counterpart across the electromagnetic
(EM) spectrum \citep{GW170817-MMA}.  These multi-messenger
observations permitted new tests of general relativity
\citep{GW170817-GRB}, measurement of the Hubble constant
\citep{GW170817-H0,GW170817-properties}, and yielded information on
neutron star physics \citep[e.g.,][]{GW170817-progenitor, Levan2017,
  Margalit2017, Bauswein2017}.

Optical follow-up observations of stellar-mass compact binary
coalescence (CBC) sources of GWs are challenging because of the large
sky localization uncertainties inherent in the LIGO-Virgo data
analysis.  With only two GW detectors, sky localizations can be
$\sim\operatorname{100--1000}\,{\rm deg}^2$
\citep{Singer2014,Berry2015}; adding additional detectors to the
network improves sky localization \citep{Veitch2012, LIGOdetect,
  Pankow2018} and enhances three-dimensional localization
\citep{Singer2016-going-the-distance,DelPozzo2018}.  Towards the end
of LIGO-Virgo's second observing run (O2) in 2017, when all three
detectors were operational, the sky localizations from LIGO and Virgo
for their triple-detector observations were
$\sim\operatorname{20--90}\,{\rm deg}^2$ \citep{GW170814detect,
  GW170817detect, GWTC-1}.  The largest cameras on 4-m and 8-m class
optical telescopes have fields of view of up to a few square degrees.
It is therefore time consuming to search thoroughly the error regions
of even the best localized GW sources, especially in the case of
binary black hole (BH-BH) mergers, for which any EM counterparts are
expected to be faint or non-existent.  Despite these challenges, early
observations of BH-BH merger sky localization error regions have been
invaluable testing grounds for optical follow-up
\citep[e.g.,][]{GW150914-EM, DES-150914, DES-151226, JGEM-151226,
  HSC-151226, Doctor2018}.  Strategies that aim to overcome the
challenges include optimizing the tiling and scheduling of wide-field
searches \citep{Coughlin2018}, and targeting the follow-up
observations on stellar mass selected galaxies located within the
three-dimensional GW localizations \citep{Nissanke2013, Hanna2014,
  Fan14, Gehrels2016}.  The latter approach was deployed to great
success in the earliest identification of the optical counterpart to
the NS-NS signal GW170817 \citep{Coulter2017}.

The luminosity distance to CBC sources is measured to
$\operatorname{30--40}$ per cent precision from LIGO-Virgo data
\citep{Berry2015, GW150914-PE, GWTC-1}.  Gravitational lensing, and in
particular strong lensing (i.e. multiple imaging), is a possible
source of systematic bias in these inferred luminosity distances,
because the amplitude of the strain signal $A$ depends on both lens
magnification $\mu$, and luminosity distance $D_{\rm L}$:
$A\propto|\mu|^{0.5}D_{\rm L}^{-1}$ (hereafter we use $\mu$ to denote
$|\mu|$).  Therefore lens magnification allows sources from greater
distances to be observed, and also means that the luminosity distance
to a lensed source inferred assuming $\mu=1$ is under-estimated by a
factor $\mu^{0.5}$ \citep{Wang1996}.  The redshift distribution of the
known galaxy and cluster strong lenses peaks close to $z=0.3$
\citep[e.g.,][]{Smith2018a}, which corresponds to a luminosity
distance of $D_{\rm L}=1.6\,{\rm Gpc}$.  For a GW source that is
initially interpreted as being located at $D_{\rm L}\,\ls\,500\,{\rm
  Mpc}$ to be reinterpreted as being strongly-lensed implies that it
must be magnified by a factor of $\mu\gs10$.  The systematic bias in
the inferred distance also means that the masses of the source, which
are calculated using the inferred source redshift \citep{Krolak1987},
are over-estimated by a factor $(1+z_{\mu=1})/(1+z)$, where
$z_{\mu=1}$ is the redshift inferred assuming $\mu=1$, and $z$ is the
true redshift of the lensed source.  Therefore, while lensing does
complicate the measurement of distance, identification of a lensed
source with LIGO-Virgo would provide a glimpse of the CBC population
at $z\,\gs\,1$, well in advance of third-generation GW detectors.

A strongly-lensed GW would travel to Earth along multiple paths
through the foreground mass concentration, and thus in principle could
be detected on more than one occasion by LIGO-Virgo, as would any EM
counterpart.  These paths differ in length, leading to a time delay
between detections of consecutive signals up to several years
\citep{Smith2018a}.  Multiple detections of a single GW source, will
create unique scientific opportunities with important advantages over
previous work and important new challenges to overcome.  The transient
nature of GW events/detections mean that EM follow-up observations
will not require expensive long-term monitoring programmes that are
typical of time delay cosmography with lensed quasars.  Moreover, the
sub-millisecond precision to which the arrival time of GW signals are
measured by LIGO-Virgo \citep{GW150914-PE, BBH-O1, GW170104detect}
will lead to a measured precision on the time delay between the
arrival of lensed GW signals that is $\gs\,8$ orders of magnitude
superior to that achievable with supernovae or quasars
\citep[e.g.,][]{Fohlmeister2007, Rodney2016}.  Therefore, in
principle, strongly-lensed GWs will yield unprecedented constraints on
the distribution of dark and luminous matter in the gravitational
lens, and a new and highly accurate measurement of the Hubble
parameter \citep{Liao2017}.  However, lens substructure and
micro-lensing may reduce the precision of such measurements
(\citealt{Suyu2018} and references therein;
\citealt{Chen2018,Tie2018}), and therefore work to address such issues
will be required.  Comparing the time delay between EM and GW images
will also enable the propagation speed of light and gravity to be
compared \citep{Collett2017,Fan17}.  Multiple detections of the same
GW source will also enable new constraints on GW polarizations,
because the number of detectors that observe the same GW signal would
grow with the number of the detections of the strongly-lensed event
\citep[cf.][]{Chatziioannou2012}.

The probability that a GW source detected to date by LIGO-Virgo is
strongly-lensed is small, because a tiny fraction of the sky is
magnified sufficiently ($\mu~\gs~10$, as discussed above) to
reinterpret the detected strain signal as originating from a source
beyond the lens population.  For example, \cite{Hilbert2008} estimate
that the source plane optical depth to $\mu>10$ for sources at $z\le2$
is $\tau_{\rm S}~\ls~10^{-5}$.  Therefore, whilst estimates for the
rate of detection of lensed GW sources vary, there is a broad
consensus that the expected rate during O1 and O2 is $\ll1\,{\rm
  yr}^{-1}$, and will rise to $\gs~1\,{\rm yr}^{-1}$ when LIGO-Virgo
reach design sensitivity in the early 2020s
\citep{Li2018,Ng2018,Smith2018a,Smith2018b}.

It will be difficult to identify that a GW source is strongly-lensed
from the LIGO-Virgo signal alone \citep{Hannuksela2019}.  This is
because the over-estimated mass of the compact objects may not appear
anomalous, and the GW sky localization uncertainties dwarf the solid
angle subtended by the strong-lensing regions of galaxies, groups, and
clusters by many orders of magnitude.  Therefore, strong evidence
beyond that available from the strain signal measured by LIGO-Virgo
will be needed to outweigh the low prior expectation that a given GW
signal was strongly-lensed.  Identification of an EM counterpart to a
GW source adjacent to the critical curve of a strong lens, and
detection of a subsequent image of the same source would provide such
evidence, and thus establish that a GW had been strongly-lensed.  This
would allow the correct source parameters to be inferred, and enable
the science outlined above.

Current observations and theoretical predictions point to galaxy
clusters dominating the optical depth to gravitationally magnifying
point sources by $\mu\ge10$.  On the observational side, all strongly
lensed images of quasars found by the Sloan Digital Sky Survey (SDSS)
that are magnified by $\mu>10$ are lensed by galaxy clusters
\citep{Inada03, Sharon2005, Oguri2010, Oguri2013, Sharon2017}.  In
contrast, individual galaxy lenses have thus far been shown to produce
only low magnification strongly lensed qusar images i.e.\ $\mu<10$
\citep{Agnello2018, Inada2005, Inada2006, Inada2007,Inada2008,
  Inada2009, Inada2014, Kayo2007, Kayo2010, McGreer2010, More2016,
  Morokuma2007, Ofek2007, Oguri2004, Oguri2005, Oguri2008, Rusu2011,
  Rusu2013}.  This picture is supported by theoretical work, notably
that of \cite{Hilbert2008}, whose optical depth to strong-lensing is
dominated by halos of mass $M_{200}>10^{13}\Msol$, i.e.\ galaxy groups
and clusters.  However, the number of highly magnified quasars seen by
SDSS is small, and \citeauthor{Hilbert2008}'s predictions pre-date
modern cosmological hydrodynamical simulations.  Therefore, more
theoretical and observational work is needed to clarify the relative
contribution of galaxy- and cluster-scale halos to high magnification
lensing of point sources like CBCs.  In Robertson et al. (in prep.),
we will consider the optical depth to strong-lensing as a function of
halo mass based on cosmological hydrodynamical simulations, and in
this article we concentrate on optical observations that explore the
strong lensing interpretation of BH-BH mergers detected by LIGO-Virgo.

We introduce a new observing strategy for identifying optical
counterparts to GW sources -- observations of strong-lensing galaxy
cluster cores located within LIGO-Virgo GW sky localization maps.  We
describe our first implementation of this strategy via rapid target of
opportunity (ToO) observations with the Gemini-South
telescope\footnote{Based on observations obtained at the Gemini
  Observatory, which is operated by the Association of Universities
  for Research in Astronomy, Inc., under a cooperative agreement with
  the NSF on behalf of the Gemini partnership: the National Science
  Foundation (United States), the National Research Council (Canada),
  CONICYT (Chile), Ministerio de Ciencia, Tecnolog\'{i}a e
  Innovaci\'{o}n Productiva (Argentina), and Minist\'{e}rio da
  Ci\^{e}ncia, Tecnologia e Inova\c{c}\~{a}o (Brazil).} and ESO's Very
Large Telescope\footnote{Based on observations made with ESO
  Telescopes at the La Silla Paranal Observatory under programme ID
  299.A-5028.} (VLT) in the nights immediately following the discovery
of the BH-BH source GW170814 \citep{GW170814detect}.  Our observations
targeted known strong-lensing galaxy clusters, selected from the list
compiled by \cite{Smith2018a}.  The main goals of our observations
were to test the feasibility of searching for optical transients in
rapid follow-up observations with small field-of-view instruments
(that are well matched to cluster cores, and not routinely used for
discovery of transient objects) on 8-m class telescopes, and to search
in earnest for candidate EM counterparts to putative strongly-lensed
GW sources.  The large aperture of Gemini-South and VLT, and the
absence of any requirement for us to explore the wider sky
localization, enabled us to conduct deep observations that are
sensitive to strongly-lensed EM counterparts down to $i=25$
independent of the actual source redshift.  Our strategy therefore
benefits from greater sensitivity than conventional searches, at the
expense of a much smaller survey volume.

We describe the details of our observing strategy, observations, and
data reduction in Section~\ref{sec:observe}, explain how we generate
difference images, and search for candidate optical counterparts, and
state our results in Section~\ref{sec:analysis}, discuss our results
in Section~\ref{sec:discussion}, and summarize in
Section~\ref{sec:conclusion}.  We assume a flat cosmology with
$H_0=67.9\,\kms\,\Mpc^{-1}$, $\Om=0.3065$ \citep{Planck15cos}.  All
celestial coordinates are stated at the J2000 epoch, and all
magnitudes are stated in the AB system.


\section{Observations}\label{sec:observe}

\subsection{Observing strategy}\label{sec:strategy}

We aim to conduct the most sensitive search to date for optical
emission from CBC sources of GWs, under the hypothesis that the
objects that we target have been strongly-lensed by a massive
foreground galaxy cluster.  We therefore select known,
spectroscopically-confirmed strong-lensing clusters located close to
the peak probability of the sky localizations of LIGO Scientific-Virgo
Collaboration (LVC) CBC alerts.  The strong-lensing regions of these
clusters span $\sim\operatorname{1--2}\,{\rm arcmin}^2$ on the sky,
and are thus perfectly matched to instruments on ground-based $8$-m
class telescopes, including the GMOS instruments on the Gemini-North
and South telescopes, and MUSE on VLT.

The most common GW sources are the coalescence of BH-BHs.  Optical
emission from BH-BHs is expected to be faint or non-existent
\citep[][and references therein]{LIGOdetect}.  Searches for optical
emission from BH-BH mergers have typically reached sensitivity limits
in the observer-frame $V/R/I$-bands of $m\,\ls\,22$ with telescopes up
to 4-m in diameter and $m\simeq\operatorname{22--24}$ with the Subaru
8-m telescope \citep[e.g.,][]{DES-150914, DES-151226, JGEM-151226,
  Arcavi2017strategy, HSC-151226, Doctor2018}.  In general, these
observations reached a sensitivity compatible with detecting a
kilonova-like counterpart to the respective BH-BH mergers and, as
discussed in Section~\ref{sec:intro}, their main aim was to implement
and test a new type of observing campaign.  We therefore adopted a
nominal goal of reaching a spectral flux density limit with GMOS and
MUSE corresponding to $i\simeq25$, in order to push the sensitivity of
EM follow-up observations in to a new regime, independent of any lens
magnification.  In particular, we note that \cite{Doctor2018} observed
the sky localization of GW170814 to a depth of $i\simeq23$ with the
Dark Energy Camera.  Moreover, after taking account of lens
magnification of (say) $\mu\simeq100$, the depth to which we observe
corresponds to a search for optical emission from BH-BHs down to
$i\simeq25+2.5\log(\mu)=30$.

Our observations are guided by the best localizations provided by the
LVC at the time.  Localizations are refined as improved analyses
become available \citep{LIGOdetect}, but since we expect a
kilonova-like optical counterpart to fade rapidly, it is not possible
to delay follow-up observations until final localizations are
communicated.  We identify the clusters for potential observations by
comparing the celestial coordinates of $130$ strong-lensing clusters
selected by \cite{Smith2018a} with the LVC sky localization.  We use
the two-dimensional sky localization to prioritize the most promising
observing targets.  Typically, we pick the strong-lensing cluster
closest to the peak of the probability distribution as the most
promising to observe.

Our ToO observing programmes at the Gemini Observatory and European
Southern Observatory (hereafter ESO) commenced in early August 2017
under programme IDs GN-2017A-DD-9, GS-2017A-DD-6, 299.A-5028
respectively.  These programmes allowed for up to $7$ epochs of
imaging observations with the GMOS instruments on the Gemini-North and
Gemini-South telescopes, and up to $3$ epochs of integral field
spectroscopy with MUSE on VLT.  The observations commence as soon as
possible after receipt of the LVC alert via a rapid ToO, and were
planned to extend over a period of one week following the alert via
regular ToO observations.

\subsection{Identification of strong-lensing clusters in the GW170814 sky localization}\label{sec:gw170814} 

\begin{figure*}
  \centerline{
    \includegraphics[width=0.4\hsize,angle=0]{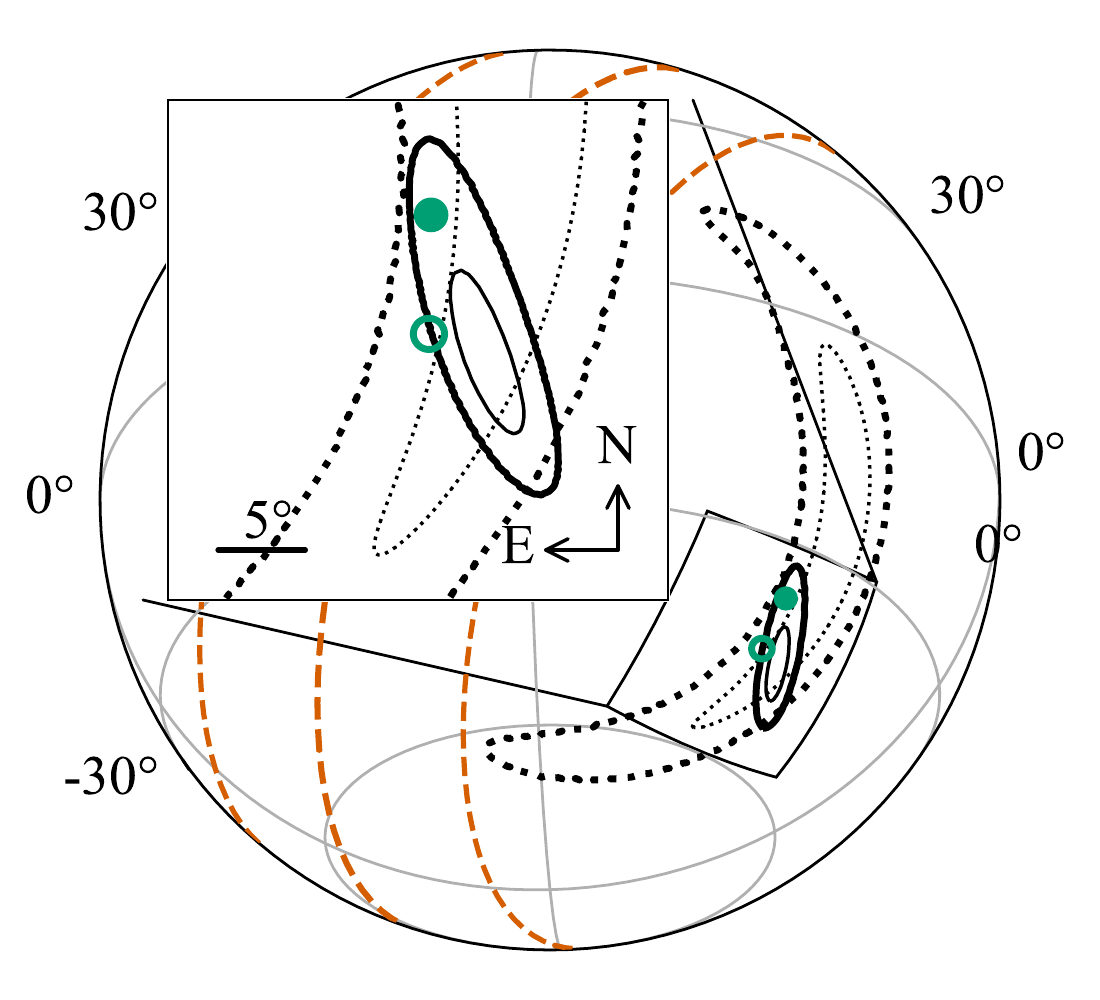} 
    \includegraphics[width=0.4\hsize,angle=0]{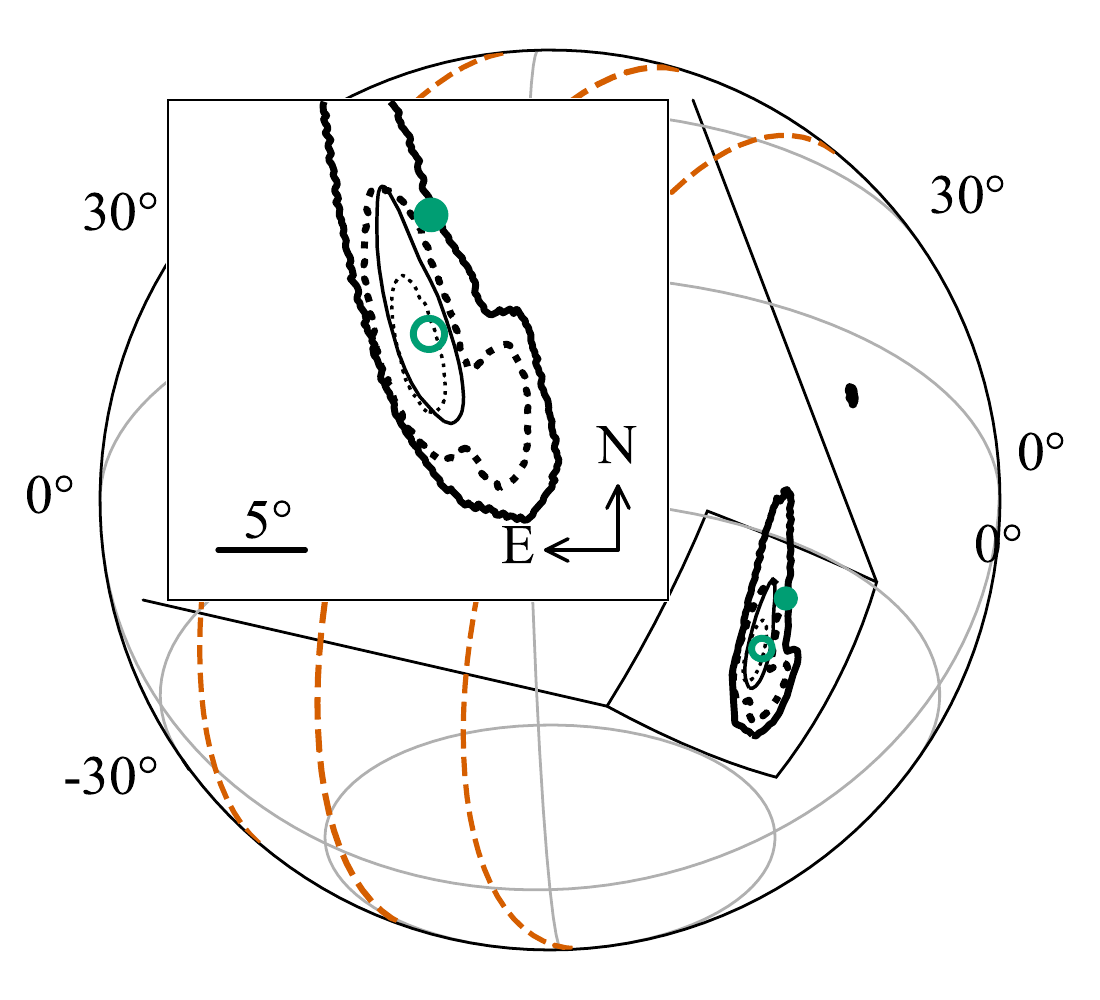}
  }
  \caption{{\sc Left} -- Initial {\sc bayestar} sky map for GW170814
    from GCN\,21474, showing the LIGO only contours (dotted), joint
    LIGO-Virgo contours (solid), Abell~3084 (filled circle), and
    SMACS\,J0304.3$-$4401 (open).  {\sc Right} -- Initial {\sc
      LALInference} sky map from GCN\,21493 (solid) and most recent
    {\sc LALInference} sky map from \citet{GWTC-1} (dotted), both
    based on LIGO-Virgo data, and the locations of both clusters.
    SMACS\,J0304.3$-$4401 (open circle) is closer to the peak of the
    sky localization than Abell\,3084 (filled) in the revised maps.
    In both panels the $90$ ($50$) per cent credible region is shown
    as the thicker (thiner) contour, and galactic latitudes of
    $\pm20^\circ$ are indicated by the dashed
    lines.} \label{fig:skyloc}
\end{figure*}

GW170814 was detected by LIGO and Virgo on August 14, 2017 at
$10{:}30{:}43$ UTC, and first announced via GCN circular on August 14,
2017 at $12{:}28{:}42$ UTC \citep{170814-Init} with an initial false
alarm rate of $\sim 1$ in $80,000$ years.  The $90$ per cent credible
region in the initial {\sc bayestar} \citep{Singer2016} sky
localization spanned $97\,{\rm deg}^2$, centered at celestial
coordinates of $(\alpha,\delta)=(02{:}44{:}00, -45{:}29{:}00)$.  We
identified one strong-lensing cluster, Abell~3084
(Table~\ref{tab:clusters}), within the $90$ per cent credible region
of the {\sc bayestar} map.  Abell~3084 lies on the contour encircling
a region within which the sky localization probability is $p=0.8$, and
which subtends $57.5\,{\rm deg}^2$ (Figure~\ref{fig:skyloc}).

As the LVC analyses were refined, the localization region evolved
(Figure~\ref{fig:skyloc}).  On August 16, 2017 at $07{:}02{:}19$ UTC,
the sky localization was updated based on the results of {\sc
  LALInference} \citep{Veitch2015}, with a revised peak close to
celestial coordinates of $(\alpha,\delta)=(03{:}06{:}00,
-44{:}36{:}00)$ and a $90$ per cent credible region spanning
$190\,{\rm deg}^2$ \citep{170814-Revised}.  We identified two
strong-lensing clusters within the $90$ per cent credible region of
the {\sc LALInference} map.  Abell~3084 lay on the contour encircling
$p=0.873$ per cent of the localization probability density,
corresponding to a region subtending $154.4\,{\rm deg}^2$ --
i.e.\ this cluster is further from the peak of the probability
distribution following the update.  SMACS\,J0304.3$-$4401
(Table~\ref{tab:clusters}) was closer to the peak of the probability
distribution, on the contour which encloses $p=0.05$ of the sky
localization probability and subtends $2.1\,{\rm deg}^2$.

When GW170814 was first announced outside of the LVC and EM follow-up
partners in October 2017 \citep{GW170814detect}, the best sky
localization peaked close to $(\alpha,\delta)=(03{:}10{:}00,
-44{:}51{:}00)$, with a $90$ per cent credible region spanning
$60\,{\rm deg}^2$, and a luminosity distance of $D_{\rm
  L}=540^{+130}_{-210}\,\Mpc$.  With this, Abell~3084 lay on the
$p=0.994$ contour that subtends $170.4\,{\rm deg}^2$, and
SMACS\,J0304.3$-$4401 lay on the $p=0.57$ contour that subtends
$16.1\,{\rm deg}^2$.  Subsequently, the LIGO-Virgo data were
recalibrated and cleaned for the O2 Catalogue results \citep{GWTC-1}.
Analysis of these improved data gave a localization peaking close to
$(\alpha,\delta)=(03{:}09{:}00, -44{:}36{:}00)$ with a $90$ per cent
credible region spanning $90\,{\rm deg}^2$, and a luminosity distance
of $D_{\rm L}=580^{+160}_{-210}\,\Mpc$.  With this, Abell~3084 lay on
the $p=0.96$ contour that subtends $170.2\,{\rm deg}^2$, and
SMACS\,J0304.3$-$4401 lay on the $p=0.30$ contour that subtends
$7.7\,{\rm deg}^2$.

\subsection{Abell~3084 and SMACS\,J0304.3$-$4401}\label{sec:clusters}

Abell~3084 and SMACS\,J0304.3$-$4401 are X-ray luminous galaxy
clusters at intermediate redshift (Table~\ref{tab:clusters}).  Virial
mass estimates of both clusters are not published to date, however
their X-ray luminosities are consistent with them both having a mass
of $M_{200}\simeq10^{15}\Msol$.  Both are spectroscopically confirmed
strong lenses, each with one multiple-image system confirmed to date.
Detailed models of the mass distribution in the cluster cores (May
2013, \citealt{Christensen2012}) are vital to interpreting the
sensitivity of our observations and any flux that we detect from a
candidate EM counterpart.  The models are most accurate for sources at
redshifts similar to the known multiple-image system redshifts; we
therefore concentrate on these redshifts when considering the
efficiency of our search for candidate EM counterparts in
Section~\ref{sec:analysis}.  \emph{Hubble Space Telescope}
({\emph{HST}}) snapshot observations with the Advanced Camera for
Surveys (ACS) are available for both clusters, through the F606W
filter (PID: 10881 and 12166).  We use these data to calibrate our
GMOS and MUSE observations (Sections~\ref{sec:gmos}~\&~\ref{sec:muse})
and in our difference image analysis (Section~\ref{sec:difference}).

\begin{table}
  \caption{Abell~3084 and SMACS\,J0304.3$-$4401}
  \label{tab:clusters}
  \begin{threeparttable}
    \begin{tabular}{p{30mm}p{18mm}p{18mm}}
      \hline 
       & Abell~3084 & SMACS\,J0304 \cr
      \hline 
      Cluster redshift & $0.22$ & $0.46$ \cr
	  \noalign{\smallskip}
	  Right ascension & $03{:}04{:}07$ & $03{:}04{:}21$ \cr
	  \noalign{\smallskip}
	  Declination & $-36{:}56{:}36$ & $-44{:}01{:}48$\cr
	  \noalign{\smallskip}
	  \raggedright $L_X[0.1-2.4{\rm keV}]$ ($10^{44}\ergs$) & \raggedright $4.0\pm0.6$\tnote{a}  & \raggedright $7.1\pm0.2$\tnote{b} \cr
	  \noalign{\smallskip}
	  \raggedright Redshift of multiple-image system & $0.764$\tnote{c} & $1.963$\tnote{d} \cr
	  \noalign{\smallskip}
	  \hline
    \end{tabular}
    \begin{tablenotes}
      \item [a] \cite{Boehringer04}
      \item [b] \cite{Repp2018}
      \item [c] May (2013) 
      \item [d] \cite{Christensen2012}
    \end{tablenotes}
  \end{threeparttable}
\end{table}

\subsection{Gemini and VLT observations}\label{sec:geminivlt}

We observed Abell~3084 with GMOS on Gemini-S on five occasions,
commencing $20$ hours after the first LVC alert pertaining to
GW170814, on August 15, 2017 UTC.  This was the first Chilean night
following the detection of GW170814.  The GMOS observations of
Abell~3084 continued after the revised sky map became available, in
order to obtain a comprehensive dataset on one cluster.  We also
observed SMACS\,J0304.3$-$4401 with GMOS on Gemini-S on two occasions
-- August 18 and 21.  The overall aim was to observe for $\sim45$
minutes on each night, with the exposure times and number of exposures
adjusted to suit the Moon phase and overhead conditions.  The
individual exposures were offset from each other randomly within a
square region of full width $30\,{\rm arcsec}$.  Observations were
performed in the $i$-band in order to minimize the impact of the Moon
on the sensitivity of the observations, and to probe rest-frame
$V$-band emission from putative lensed GWs at $z\,\gs\,0.5$.

We triggered a rapid ToO on VLT with MUSE within $3$ hours after LVC
announced the detection of GW170814.  This observation was executed at
the telescope on August 17, 2017 UTC, and repeated on August 20, 2017
UTC.  The delay between our trigger and the first MUSE observation was
due to a combination of visitor mode time and engineering time on the
intervening nights.  Each observation comprised three exposures of
duration $980\,{\rm s}$ and spanned the wavelength range $475\,{\rm
  nm}<\lambda<930\,{\rm nm}$.

All of the GMOS and MUSE data were obtained at high elevation and with
excellent seeing of $0.5\,{\rm arcsec}\le{\rm FWHM}\le1\,{\rm
  arcsec}$.  Details of the GMOS and MUSE observations are listed in
Table~\ref{tab:obs}, and the reduction of GMOS and MUSE data are
described in Sections~\ref{sec:gmos}~\&~\ref{sec:muse}, respectively.

\begin{table*}
\caption{Follow-up observations of strong-lensing clusters within sky
  localization of GW170814 }
\label{tab:obs}
\centering
\begin{threeparttable}
\begin{tabular}{cccccc}
      \hline 
      Visit & Start of observation (UTC) & Airmass\tnote{a}    & Integration & Seeing\tnote{b}   &  Sensitivity\tnote{c} \cr 
            &                            &       & time (ks)   & (arcsec) & \cr

      \hline 
      \multispan6{\sc \hfil GMOS Observations of Abell~3084\hfil}\cr
      \hline
      1 & August 15, 2017, $06{:}40{:}40$ & $1.29$ & $2.9$ & $0.72$ & $24.7$ \cr
2 & August 17, 2017, $07{:}50{:}52$ & $1.09$ & $2.3$ & $0.49$ & $25.2$ \cr 
      3 & August 18, 2017, $06{:}14{:}13$ & $1.33$ & $3.1$ & $0.81$ & $24.8$\cr 
      4 & August 21, 2017, $07{:}09{:}38$ & $1.05$ & $3.1$ & $0.76$ & $24.9$ \cr 
      5 & August 28, 2017, $07{:}44{:}48$ & $1.03$ & $3.1$ & $0.84$ & $24.9$ \cr
      \hline
      \multispan6{\sc \hfil GMOS Observations of SMACS\,J0304.3$-$4401\hfil}\cr
      \hline
      1 & August 18, 2017, $07{:}23{:}35$ & $1.14$ & $2.3$ & $1.01$ & $24.9$ \cr
      2 & August 21, 2017, $05{:}58{:}47$ & $1.35$ & $3.1$ & $0.88$ & $25.1$\cr 
      3 & August 27, 2017, $08{:}02{:}50$ & $1.04$ & $2.9$ & $0.97$ & $25.0$\cr 
      \hline
      \multispan6{\sc \hfil MUSE Observations of Abell~3084\hfil}\cr
      \hline
      1 & August 17, 2017, $08{:}02{:}13$ & $1.10$ & $2.9$ & $0.85$ & $25.8$ \cr 
      2 & August 19, 2017, $07{:}46{:}50$ & $1.12$ & $2.9$ & $0.82$ & $25.9$ \cr 
      \hline
    \end{tabular}
    \begin{tablenotes}
    \item [a] The airmass at the mid-point of the observation.
    \item [b] Mean full width at half maximum of point sources in the
      reduced data, with a typical error on the mean of
      $\sim0.02\,{\rm arcsec}$.
    \item [c] $5\sigma$ point source sensitivity within a photometric
      aperture of diameter $2\,{\rm arcsec}$, estimated from the
      magnitude at which the median photometric uncertainty is
      $0.2\,{\rm magnitudes}$.  The sensitivity of the MUSE
      observations is stated in the F606W-band.
    \end{tablenotes}
\end{threeparttable}
\end{table*}

\subsection{GMOS data reduction}\label{sec:gmos}

Individual GMOS exposures were de-biased, dark-subtracted,
flat-fielded, and sky-subtracted in the standard manner using the {\sc
  gemini} package in {\sc iraf}, to produce both a single science
frame comprising the mosaiced individual chips, and a bad-pixel map,
for each exposure.  The bad-pixel maps were applied to the science
frames and the individual masked science frames were then combined in
to a single stacked frame per visit using the {\sc imcombine} task in
{\sc iraf}.  The full width at half maximum of point sources in the
reduced frames is consistently sub-arcsecond (Table~\ref{tab:obs}).
The reduced and stacked frames were aligned with the first visit for
that target to a typical root-mean-square residual accuracy of
$0.03\,{\rm pixels}$ using the {\sc iraf} tasks {\sc geomap} and {\sc
  geotran}.

\begin{figure}
\centerline{\includegraphics[width=\columnwidth]{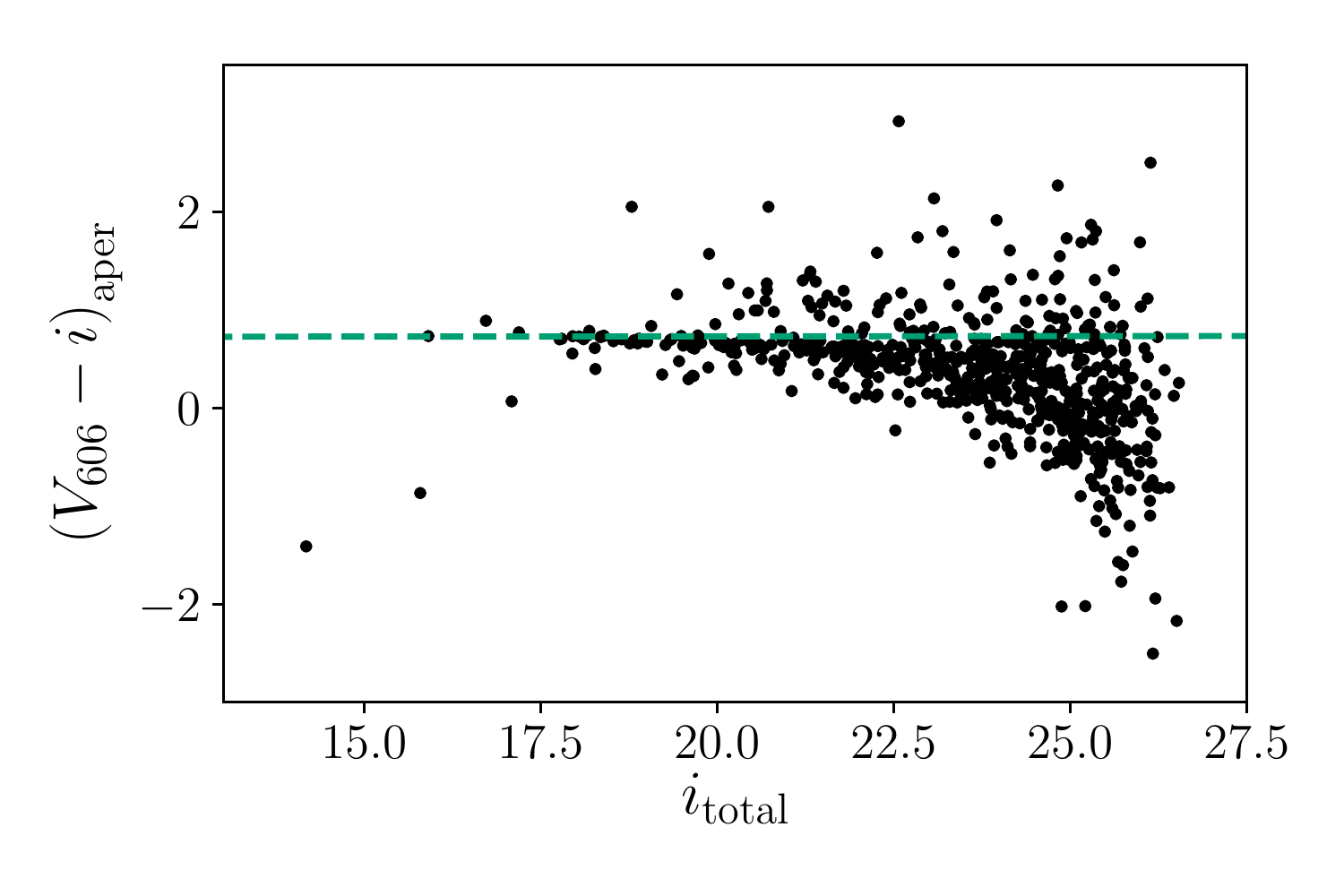}}
  \caption{Example colour-magnitude diagram for Abell~3084 based on
    our GMOS and archival \emph{HST}/ACS data.  The horizontal dashed
    line shows the predicted colour of a single stellar population
    that formed at high redshift and evolved passively to the cluster
    redshift ($z=0.22$).  The photometric zero point of the GMOS frame
    was chosen to match the colour of the sequence of massive ($i<20$)
    cluster early-type galaxies to this prediction.}
  \label{fig:colmag}
\end{figure}

We searched the available USNO and GSC catalogues for sources of well
calibrated $i$-band magnitude within the field of view of our GMOS
data.  The size and depth of the GMOS imaging even in short exposures
meant that there was no overlap between unsaturated bright stars as
seen by Gemini and faint stars measured in all-sky surveys.  We
therefore calibrated the GMOS frames by measuring the $(V_{606}-i)$
colours of sources detected in both the archival \emph{HST}/ACS data
and our GMOS data, and selecting the photometric zero point that
yields the correct colours for massive early-type galaxies in
Abell~3084 and SMACS\,J0304.3$-$4401 respectively.  These colours were
computed using the {\sc EzGal}
code,\footnote{\href{http://www.baryons.org/ezgal/}{www.baryons.org/ezgal/}}
using a single stellar population that formed at high redshift and
evolved passively to the relevant cluster redshifts based on the
\citet{Bruzual03} population-synthesis code.  The predicted colours
are insensitive to the fine details of how we choose the formation
redshift and the metallicity.  We show an example $(V_{606}-i)/i$
colour-magnitude diagram for one of the observations of Abell~3084 in
Figure~\ref{fig:colmag}.

\subsection{MUSE data reduction}\label{sec:muse}

The MUSE observations were reduced using version 2.0 of the data
reduction software \citep{Weilbacher2015}.  The process includes basic
calibration (bias removal, flat-fielding, wavelength and geometrical
calibration) and the production of datacubes for each exposure
following sky subtraction, flux calibration and telluric correction.
These datacubes were matched in astrometry to the relevant \emph{HST}
observation discussed in Section~\ref{sec:clusters}.  In all cases,
only a constant offset in right ascension and declination was applied,
as no significant rotation was found.  The measured offsets and
rotation were applied back to the original list of pixels (the
pixel table) so that the datacubes can be produced in a
single interpolation step to limit the effect on the noise properties.
Each exposure datacube was treated for sky subtraction residuals using
the PCA method implemented in the ZAP v2.0 software \citep{ZAP}.  We
then combined all zapped exposures taken during each of the two
observations.

We assessed the image quality of the MUSE observations by performing a
Moffat spatial fit of the bright unsaturated stars in each exposure.
The seeing was stable in each night with average values provided in
Table \ref{tab:obs}.  We also estimated the cloud extinction to be
just a few per cent and thus negligible, based on a comparison between
MUSE pseudo-F606W frames (see Section~\ref{sec:difference})
and relevant \emph{HST} frames.


\section{Analysis and results}\label{sec:analysis}

\subsection{Difference images}\label{sec:difference}

\begin{figure*}
  \centerline{
    \includegraphics[width=\hsize,angle=0]{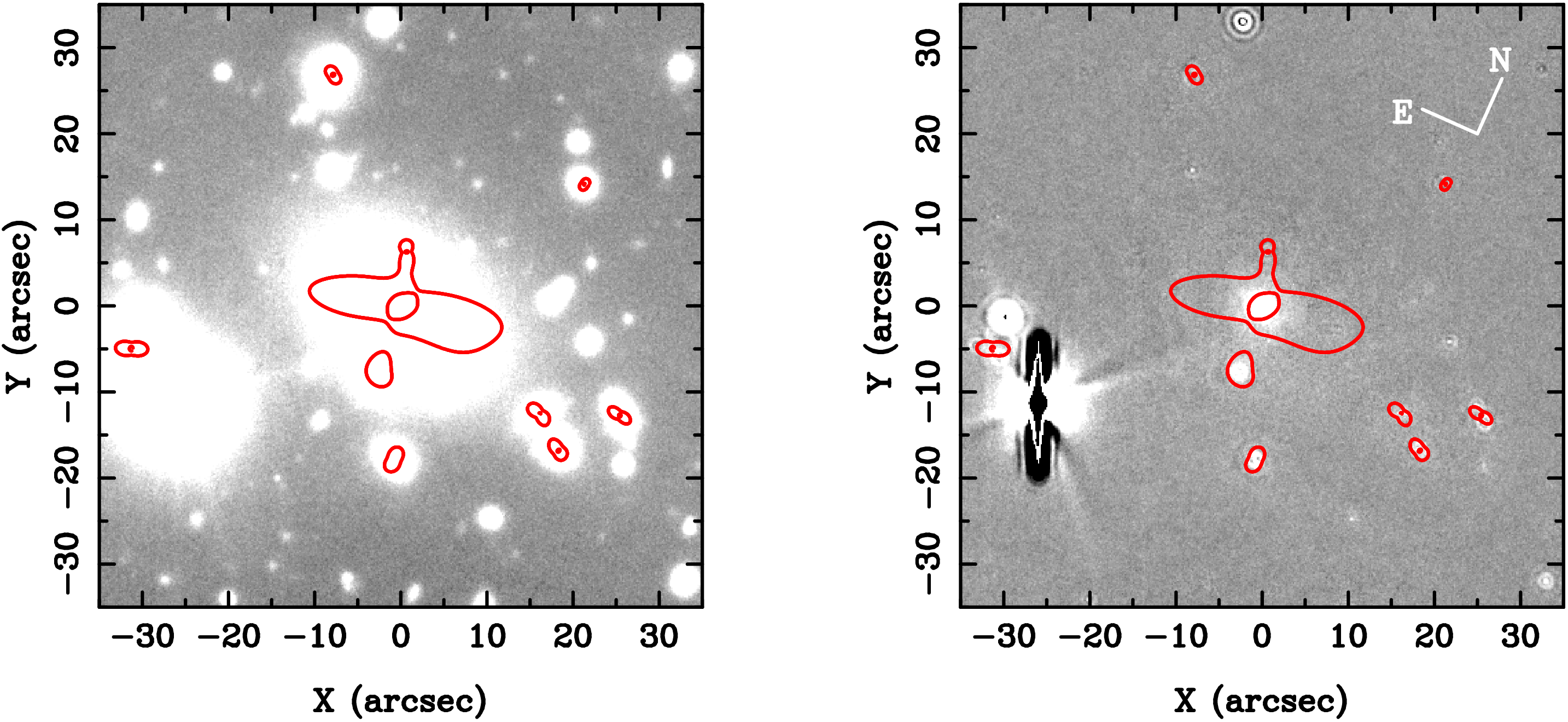}
  }
  \vspace{10mm}
  \centerline{
    \includegraphics[width=\hsize,angle=0]{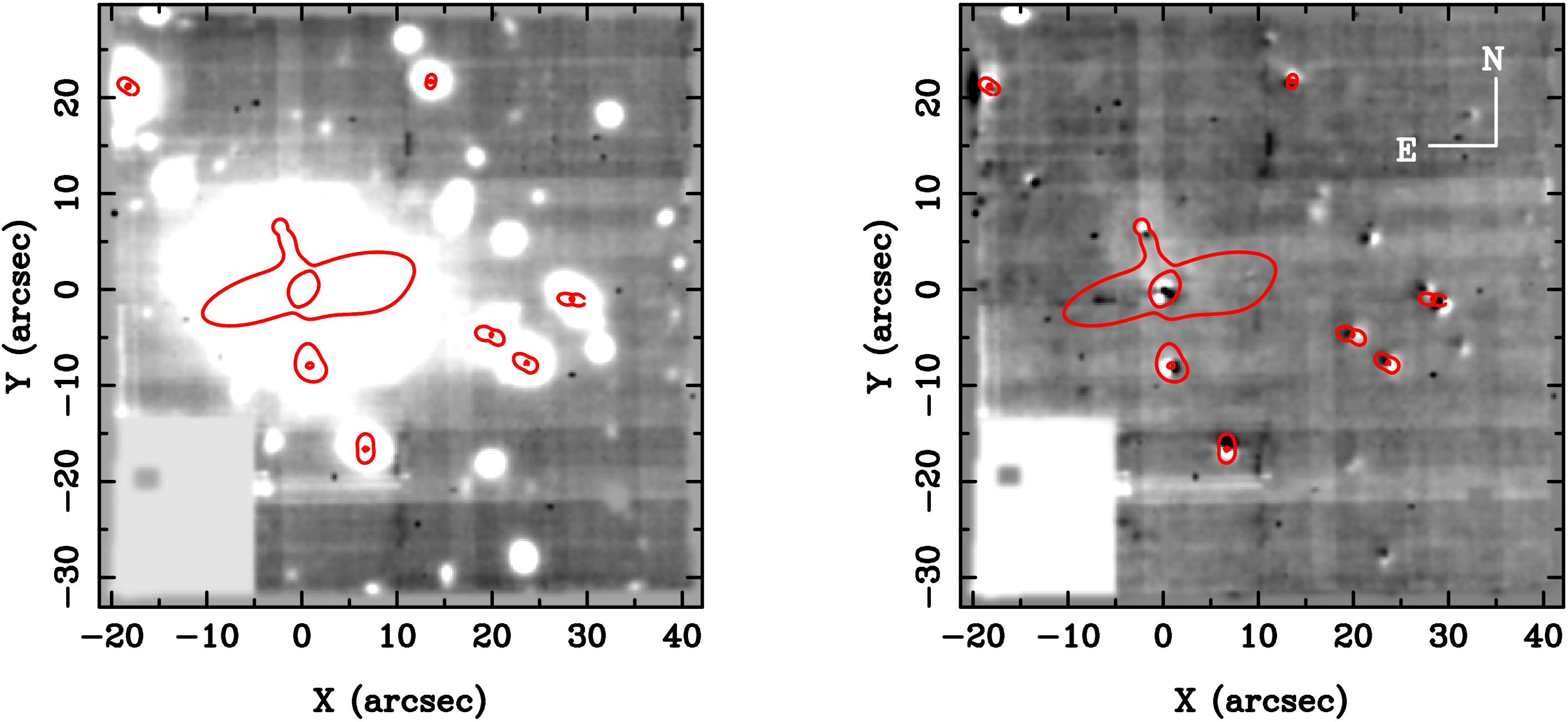}
  }
  \vspace{5mm}
  \caption{Central region of Abell~3084, with the brightest cluster
    galaxy at the origin in each panel.  Examples from our data are
    shown as follows: GMOS $i$-band observation (top left); GMOS-based
    difference image (top right); MUSE observation convolved with the
    F606W filter (bottom left); MUSE/\emph{HST}-based difference image
    (bottom right).  The red curves show the critical curves for
    $\zs=0.764$. }
  \label{fig:a3084}
\end{figure*}

\begin{figure*}
  \centerline{
    \includegraphics[width=\hsize,angle=0]{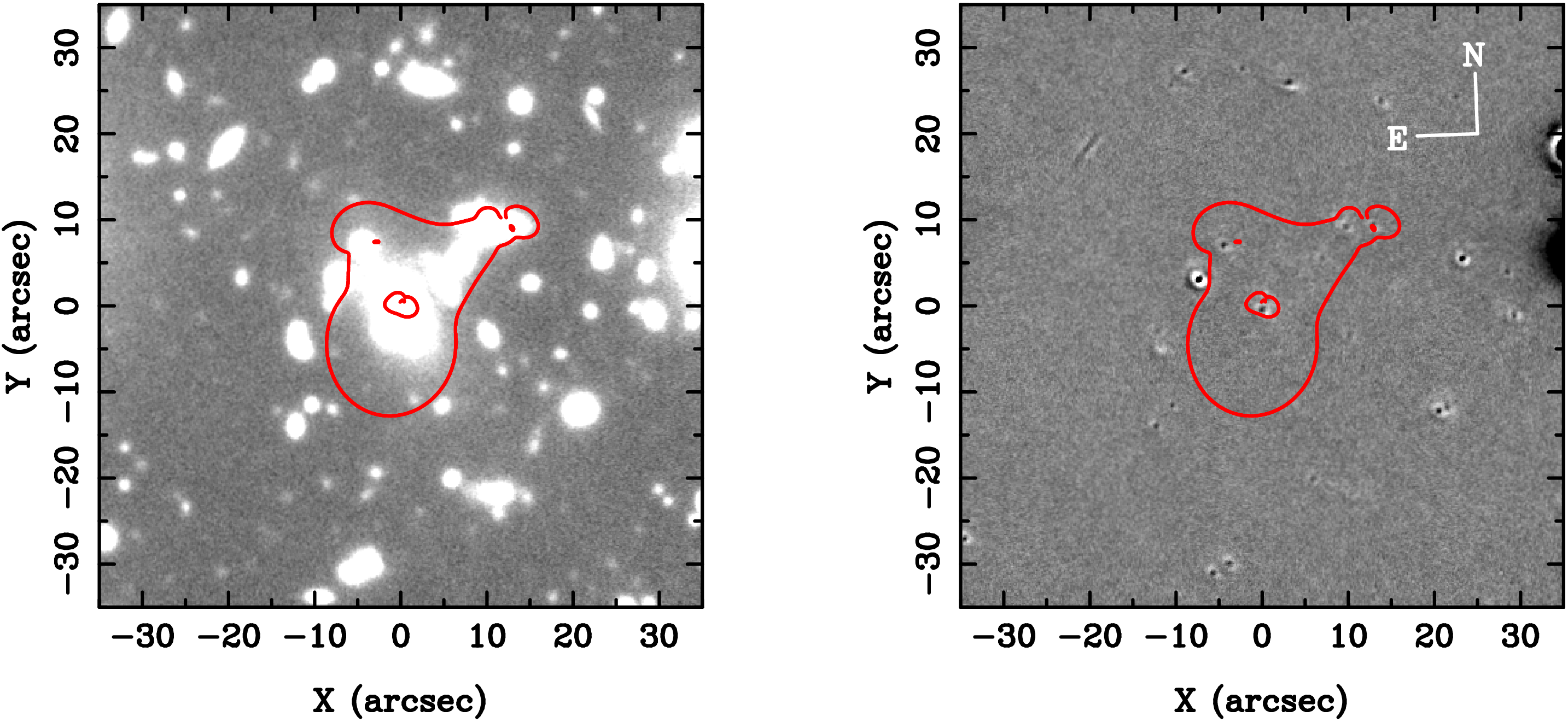}
  }
  \vspace{5mm}
  \caption{Central region of SMACS\,J0304.3$-$4401, centred on the
    brightest cluster galaxy.  Examples from our data are shown as
    follows: GMOS $i$-band observation (left); GMOS-based difference
    image (right).  The red curve is the tangential critical curve for
    $\zs=1.963$.}
  \label{fig:smacs}
\end{figure*}

We normalised each reduced and stacked GMOS frame to an exposure time
of one second, and then matched the image quality between visits using
the {\sc psfmatch} task in {\sc iraf}.  The point spread function
(PSF) model was empirical and based on $10$ isolated, unsaturated,
high signal-to-noise ratio stars in the vicinity of the strong-lensing
region of each frame.  These matched frames were then subtracted from
each other in pairs to produce difference images, examples of which
are shown in Figures~\ref{fig:a3084}~\&~\ref{fig:smacs}.

\begin{figure}
  \centerline{
    \includegraphics[width=0.9\hsize,angle=0]{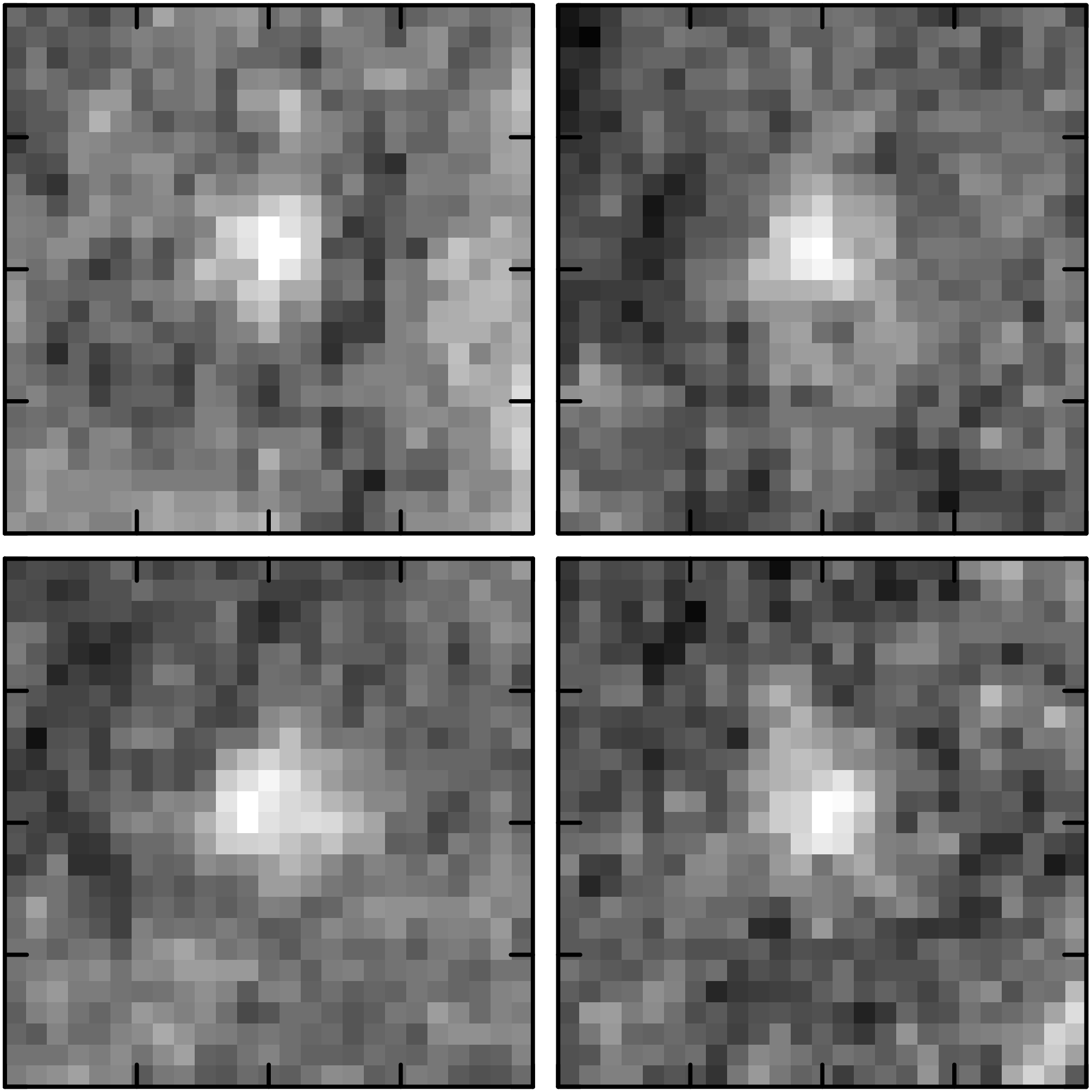}
  }
  \vspace{5mm}
  \caption{Example fake $i=25$ point sources that were injected in to
    the GMOS observations of Abell~3084.  Each panel is
    $4\times4\,{\rm arcsec}$.}
  \label{fig:fake}
\end{figure}

The wavelength range of the MUSE datacube ($475<\lambda<930\,{\rm
  nm}$) obtained from our observations of Abell~3084 enables great
freedom in the choice of reference image when constructing a
difference image.  We selected the \emph{HST}/ACS observation through
the F606W filter (Section~\ref{sec:clusters}) as the reference image.
We therefore created a F606W pseudo-image from the MUSE datacube for
each night upon which we observed Abell~3084, by multiplying the
datacubes with the transmission curve of the F606W filter, and
integrating under the transmission curve.  The difference images were
then produced by minimizing residuals in all detected sources,
following a similar procedure described by \citet{Bacon2017} in the
MUSE observations of the Ultra Deep Field.  This optimization is
performed over: the Moffat model of the MUSE PSF, a possible
astrometric shift between the two images, and the sky background level
and average sky transmission.  An example difference image is shown
in Figure~\ref{fig:a3084}.

\subsection{Searching for transients in imaging data}\label{sec:transients}

We searched the difference images for transient sources both manually
and automatically.  The manual search was performed by three authors
(GPS, MB, AR), and identified no sources consistent with being a point
source in the difference images.  The only sources found were either
residuals from the PSF matching close to the centres of bright
galaxies, artefacts related to saturated stars, or a small number of
residual bad pixels and cosmic rays.  The automated search was
performed with {\sc SExtractor} \citep{Bertin96}, and found no sources
other than PSF matching residuals and artefacts relating to saturated
stars.

To calibrate the sensitivity of our GMOS search for candidate
transients, we injected fake transients in to the data in the
following manner, and attempted to recover them both manually and
automatically.  We cut out $31\times31\,{\rm pixel}$ stamps from the
GMOS data around the same, unsaturated, high signal-to-noise, stars
used for the image quality matching.  Each injected source was
randomly chosen to be one of these bright star stamps, with the flux
scaled accordingly, and shot noise added.  To test the efficacy of
manual recovery, we generated synthetic data sets, each containing
(randomly and unknown to those searching) between zero and ten
injected sources, with $i$-band magnitudes drawn uniformly from the
range $21<i<27$.  All injected sources with $i<25$ were identified by
at least two searchers, while no sources with $i>25$ were identified
by anyone.  These results were independent of whether sources were
injected randomly within $20\,{\rm arcsec}$ of the brightest cluster
galaxy (BCG), or into faint background galaxies (including the
spectroscopically confirmed multiply-imaged galaxies) following the
methods described by \citet{2010ApJ...718..876S}.

\begin{figure}
  \includegraphics[width=\columnwidth]{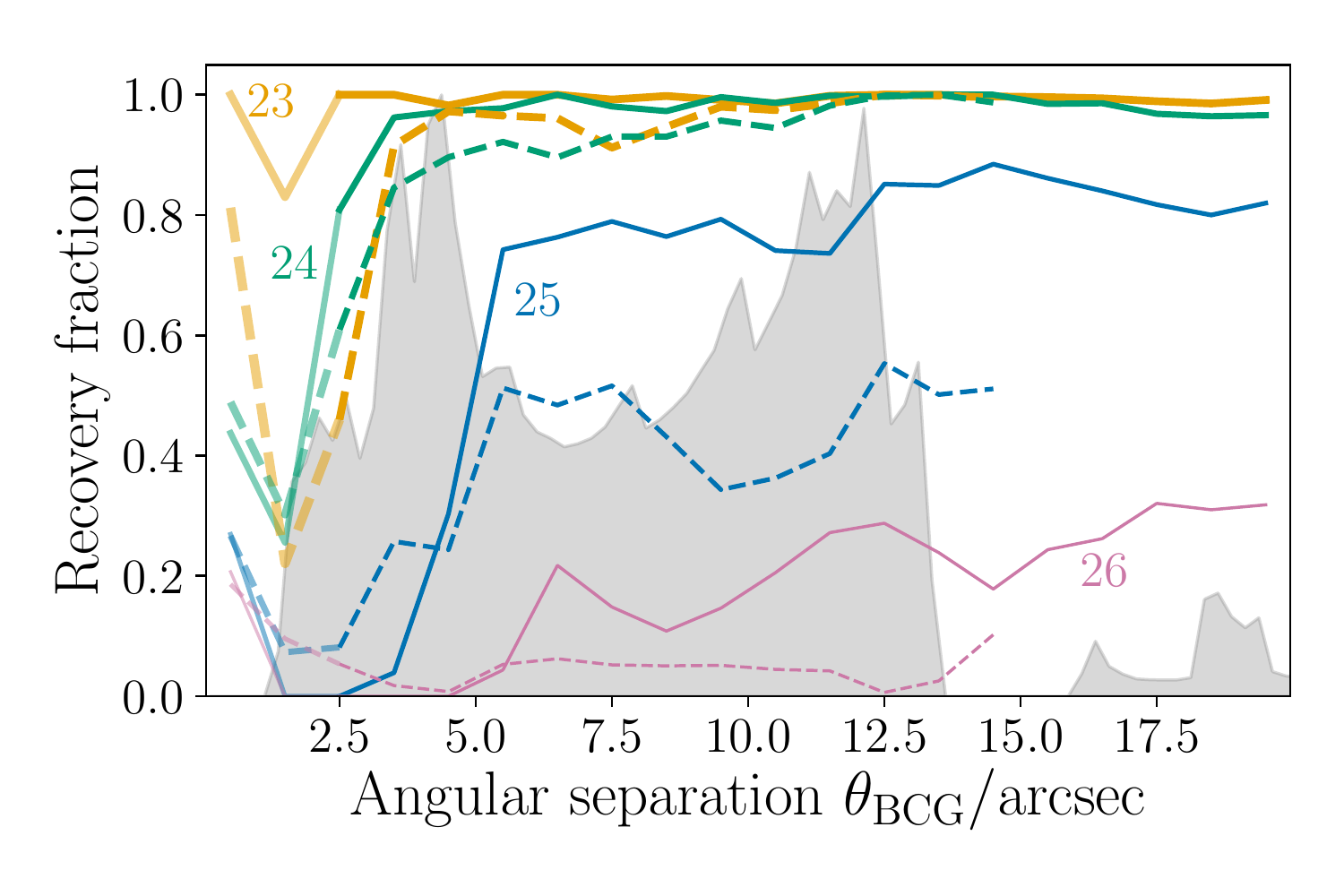}
  \caption{The recovery fraction of injected sources with different
    $i$-band magnitudes (as labeled), as a function of angular
    separation from the BCG.  The solid lines show fractions for
    sources injected into GMOS observations of Abell~3084, while the
    dashed lines are for sources injected into the MUSE/\emph{HST}
    data for the same cluster.  The lines are faded at small radii
    where the recovery of injected sources is confused by the presence
    of structured residuals near the BCG centre.  The radial
    distribution of image-plane solid angle magnified sufficiently to
    reinterpret the LIGO-Virgo detection of GW170814 as originating
    from $z=0.764$ is shown by the grey area. }
  \label{fig:fake_source_recovery}
\end{figure}

To test the automated search, we generated $10^4$ synthetic datasets
comprising sources injected at each integer magnitude in the range
$21\le i\le26$.  {\sc SExtractor} was then run on these synthetic data
sets, with the source being detected if {\sc SExtractor} finds a
source within $0.5\,{\rm arcsec}$ of the injected source location.  At
small angular separations from the BCG ($\theta_{\rm BCG}$) the
probability of sources being detected is low due to the residuals in
the difference images (Figure~\ref{fig:fake_source_recovery}).  The
source recovery rate at the very centre (within $<0.5\,{\rm arcsec}$
of the BCG) is artificially boosted because {\sc SExtractor} detects
the residuals at the centre of the BCG (in the absence of an injected
source) as a source.  However, these scales are generally smaller than
the high magnification region close to the radial critical curve, and
therefore the residuals at the centre of the BCG are not a major
limiting factor in our analysis.

Sources at the typical $5\sigma$ point source sensitivity of our GMOS
data ($i\simeq25$) are detected with the expected $\sim80$ per cent
completeness at $\theta_{\rm BCG}\,\gs\,5\,{\rm arcsec}$, and sources
at $i\simeq24$ are detectable right down to angular separations of a
few arcseconds.  We put this in to context by plotting in grey in
Figure~\ref{fig:fake_source_recovery} the solid angle that is
magnified by the amount required to reinterpret the strain signal from
GW170814 as coming from $z=0.764$, which is the redshift at which the
lens model of Abell~3084 is most robust (Section~\ref{sec:clusters}),
also taking in to account the uncertainty on the luminosity distance
to GW170814.  Use of $z=0.764$ is by way of example only, and
motivated by not wishing to extrapolate the lens model to redshifts at
which it is not constrained. The peaks at $\theta_{\rm
  BCG}\simeq4\,{\rm arcsec}$ and $\theta_{\rm BCG}\simeq12\,{\rm
  arcsec}$ correspond to the radial and tangential critical curves
respectively (see the inner and outer red curves in
Figure~\ref{fig:a3084}).  Our search is therefore sensitive at $>80$
per cent completeness to point sources at $i\,\ls\,24$ that are
adjacent to either critical curve and to point sources at $i\,\ls\,25$
that are adjacent to the tangential critical curve.

We performed similar manual and automated tests with MUSE images,
injecting sources into the inner $15\,{\rm arcsec}$ of the Abell~3084
MUSE/\emph{HST} difference image, obtaining similar results to the
GMOS tests described above.  As a consequence of the smaller field of
view of MUSE compared to GMOS, we did not have a large number of
bright stars to use as templates for point sources, but instead used
the best-fit Moffat profile discussed in Sec.~\ref{sec:muse}.  We
interpret the slight shortfall in the fraction of fake sources that
are recovered from the MUSE/\emph{HST} difference images relative to
the GMOS images as being caused by the flux sensitivity of the short
\emph{HST} observation being inferior to that of the MUSE
observations, after smoothing the \emph{HST} data to match the
ground-based seeing.

\subsection{Searching for transients in MUSE data cubes}\label{sec:transmuse}

\begin{figure}
  \includegraphics[width=\columnwidth]{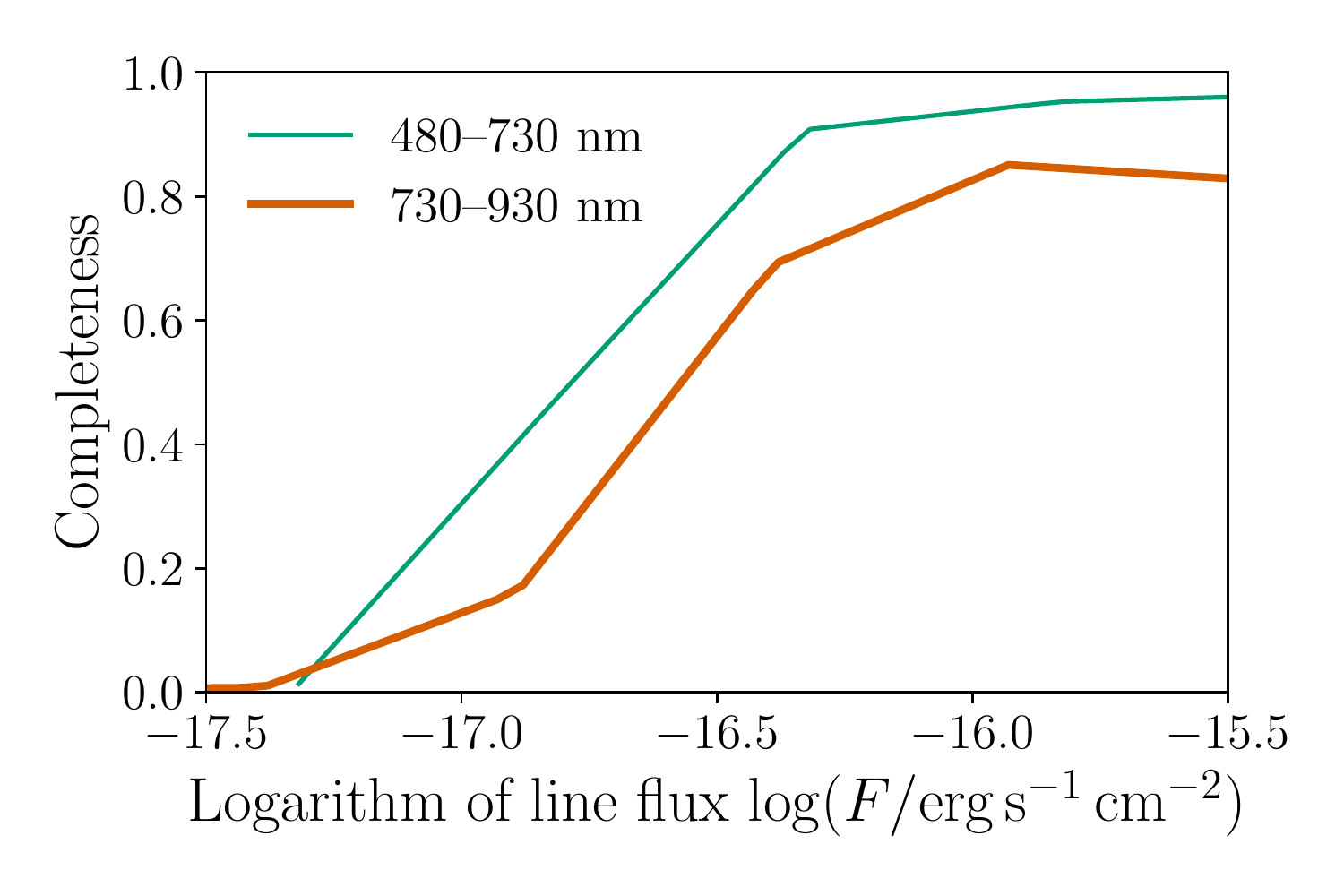}
  \caption{Emission line sensitivity in our search for transients in
    the individual (per night) MUSE datacubes.  Presented is the
    average completeness level of recovering emission line point
    sources with {\sc muselet} as a function of line flux, which
    slightly varies across the cube as a function of the wavelength
    range.}
  \label{fig:muse_linesens}
\end{figure}

In addition to the search for continuum sources based on pseudo-images
created from the MUSE datacubes, we have performed an automatic search
for emission lines across all wavelengths.  This is done using the
{\sc muselet} detection software, which has been used in the past to
automatically search for line emitters in MUSE blank fields
\citep{Drake2017a,Drake2017b} as well as lensing cluster fields
\citep{Mahler2018,Lagattuta2017}.  {\sc muselet} is publicly available
as part of the MUSE Python Data Analysis Framework
\citep[MPDAF,][]{Conseil2016} software suite.  It is a {\sc
  SExtractor}-based detection tool based on a continuum-subtracted
datacube where each wavelength plane is replaced by its corresponding
narrow-band image, optimised for the detection of typical ${\rm
  FWHM}=150\,\kms$ line width line sources.  The sensitivity of {\sc
  muselet} in recovering point-source line emitters has been
extensively tested by \citet{Drake2017b} in the MUSE Ultra Deep Field
\citep{Bacon2017}, and we rescale their results to the exposure times
of individual datacubes in each night.  The values show that our
search is complete for line fluxes brighter than
$5\times10^{-17}\,{\rm erg\,s^{-1}\,cm^{-2}}$
(Figure~\ref{fig:muse_linesens}).

Removing obvious false detections (sharper than the spatial and/or
spectral PSF) through visual inspection, we compare the detection of
line emitters found in each datacube as well as the combination of
both nights with the reference \emph{HST} image.  The only sources
appearing as line emitters in MUSE and absent from \emph{HST} are
clear Lyman-alpha emitters with strong equivalent width identified
from the shape of their spectral lines and/or as multiple images.

\section{Discussion}\label{sec:discussion}

We now discuss the empirical sensitivity of our observations of
strong-lensing cluster cores in the context of the plausible EM
counterparts to CBC GW sources.  We first discuss the cancellation of
the inverse square law by lens magnification for strongly-lensed point
source EM counterparts to GWs (Section~\ref{sec:cancel}), and then
apply this to kilonova-like EM counterparts
(Section~\ref{sec:kilonova}), and EM counterparts that might resemble
emission from known low mass X-ray binaries within the Milky Way
during outburst (Section~\ref{sec:accretion}).

\subsection{The inverse square law and lens magnification}\label{sec:cancel}

Consider a hypothetical GW source that is inferred to be at a
luminosity distance $D_{{\rm L},\mu=1}$ assuming $\mu=1$, and that has
an EM counterpart of absolute magnitude $M$ in an arbitrary pass-band.
If this is actually a distant source at $D_{\rm L,true}$ that is
gravitationally magnified by $\mu>1$, then its apparent magnitude in a
pass-band that probes the same rest-frame wavelength range as the
pass-band relevant to $M$ (or, equivalently, assuming $k$-corrections
are negligible), is given by:
\begin{equation}
  m=M+5\log(D_{\rm L,true})-5-2.5\log(\mu),
  \label{eqn:apparent}
\end{equation}
with $D_{\rm L,true}$ expressed in units of parsecs.  As discussed in
Section~\ref{sec:intro}, the lens magnification that appears in
Equation~\eqref{eqn:apparent} is given by:
\begin{equation}
  \mu=\left(\frac{D_{\rm L,true}}{D_{{\rm L},\mu=1}}\right)^2.
  \label{eqn:mu}
\end{equation}
The apparent magnitude of the EM counterpart of a lensed GW therefore
depends only on the absolute magnitude of the counterpart and the
luminosity distance at which the source is placed when (incorrectly)
assuming $\mu=1$:
\begin{equation}
  m=M+5\log(D_{{\rm L},\mu=1})-5.
  \label{eqn:frozen}
\end{equation}
The apparent magnitude of a lensed EM counterpart is therefore set by
the initial analysis of the signal detected by LIGO-Virgo assuming
$\mu=1$.  Therefore observations capable of detecting a given EM
counterpart located at $D_{\rm L,\mu=1}$ are also able to detect the
same type of counterpart if it is lensed, independent of the true
redshift.

\subsection{Kilonovae}\label{sec:kilonova}

Optical follow-up observations of BH-BH sources have so far generally
aimed to achieve the sensitivity required to detect a kilonova at
$D_{\rm L,\mu=1}$ \citep[e.g.,][]{DES-150914, DES-151226, JGEM-151226,
  Arcavi2017strategy, HSC-151226, Doctor2018}.  We therefore consider
how bright a kilonova-like counterpart to GW170814 would have been if
it had been lensed.  The kilonova counterpart to GW170817 had an
absolute magnitude in optical bands of $M\,\ls\,{-13.5}$ in the few
days after discovery \citep[e.g.,][]{Villar2017,Arcavi2018}.  We
therefore adopt $M\simeq-13.5$ in the rest-frame $BVR$-bands as a
template for our calculations.  If GW170814 was strongly-lensed, and
had a kilonova-like EM counterpart, then based on its inferred
luminosity distance, {\bf $D_{{\rm L},\mu=1}=580^{+160}_{-210}\,\Mpc$}
(Section~\ref{sec:gw170814}), it would have an apparent magnitude of
$m\simeq\operatorname{24--26}$ {\bf in the rest-frame $BVR$-bands}
independent of its luminosity distance.  Our observations reach a
depth of $m\simeq\operatorname{25--26}$ (Table~\ref{tab:obs}), with a
sensitivity to transient sources in our difference image analysis of
$m\simeq\operatorname{24--25}$ at $\gs\,80\%$ completeness.  These
sensitivities are in the observer-frame $V_{606}-$ and $i$-bands, and
can thus be compared with the rest-frame bands quoted above for
sources at $z\simeq1$.  It should therefore have been possible to
detect a kilonova-like EM counterpart if GW170814 was strongly-lensed
by one of the clusters that we observed.

A kilonova-like EM counterpart to GW170814 is more plausible if
GW170814 actually comprised one or more NS in a NS-NS or NS-BH system.
The primary means of identifying a GW source as a NS-NS, NS-BH or
BH-BH is from the inferred component masses, as NSs have a maximum
mass of $<\,3\,\Msol$
\citep{Rhoades1974,Kalogera1996,Margalit2017}.\footnote{Neutron stars
  could also be disambiguated from the imprint of tidal effects on the
  signal. These are difficult to measure, and could not be
  conclusively identified for GW170817 despite its high
  signal-to-noise ratio, without assuming that the source comprised
  neutron stars \citep{GW170817detect,GW170817-properties}.}  For a
particular GW signal, the inferred rest-frame mass of the compact
objects is inversely proportional to $(1+z)$ where $z$ is the redshift
of the GW source.  Therefore, the rest-frame mass of GW sources would
be revised downwards from the initial estimate by LIGO-Virgo, if they
are subsequently identified as being strongly-lensed.  The source
frame masses of the individual BHs that comprise GW170814 are
$30.7^{+5.7}_{-3.0}\,\Msol$ and $25.3^{+2.9}_{-4.1}\,\Msol$
\citep{GWTC-1}, which correspond to $\sim34\,\Msol$ and
$\sim28\,\Msol$ in the detector frame, i.e.\ $\gs\,9$ times larger
than the maximum NS mass.  Therefore, if GW170814 were
strongly-lensed, then it would have to be at $z\,\gs\,8$ for one or
both of the two BHs to be reinterpreted as a NS.  This implies a lens
magnification of $\mu\gs10^4$, which is possible for point sources
located behind galaxy cluster lenses \citep{MiraldaEscude1991,
  Diego2018, Kelly2018, Rodney2018}.  Our observations and difference
image analysis yielded no transient sources in the strong-lensing
regions of Abell~3084 and SMACS\,J0304.3$-$4401 down to $m=25$.
Therefore, if we assume that the rest-frame ultraviolet luminosity of
a kilonova-like EM counterpart is similar to the rest-frame optical
luminosity of GW170817, we can exclude the interpretation of GW170817
as a NS-NS or NS-BH merger at $z>8$ that has been strongly lensed by
either of these clusters.

More generally, our analysis shows that the cancellation of the
inverse square law by the degeneracy between luminosity distance and
lens magnification has important implications for the EM follow up of
GW sources.  Specifically, that the detection of EM counterparts to
lensed NS-NS and NS-BH sources are within the reach of deep ground-based
optical observations with $8$-m class telescopes.  Our difference
image analysis is sensitive to transients as faint as $i\simeq25$ and
are thus sensitive to kilonova-like counterparts to lensed sources
that are initially identified at $D_{{\rm L},\mu=1}\simeq500\,\Mpc$,
independent of their true distance.

In the future, it may be more fruitful to search for lensed optical
kilonova-like counterparts to GW sources initially identified as
low-mass BH-BH systems, with individual BH masses of $<10\,\Msol$,
because such lower masses imply a less extreme lens magnification.
For example, GW170608 \citep{GW170608detect} would have been an ideal
target for our observing programme, if we had commenced our observing
programmes at VLT and Gemini just a few months earlier.  GW170608
comprised (assuming $\mu=1$) two BHs of masses
$10.9^{+5.3}_{-1.7}\,\Msol$ and $7.6^{+1.3}_{-2.1}\,\Msol$
respectively, at $z=0.07^{+0.02}_{-0.02}$, which corresponds to
$D_{{\rm L},\mu=1}=320_{-110}^{+120}\,\Mpc$ \cite{GWTC-1}.  Following
the reasoning outlined above, it is possible that this is a lensed
NS-NS source at $z\simeq\operatorname{2--5}$, that is magnified by
$\mu\simeq\operatorname{10^3--10^4}$.  This level of magnification is
similar to that suffered by the strongly-lensed individual blue giant
star dubbed ``Icarus'' \citep{Kelly2018}.  A kilonova-like counterpart
to GW170608 would have an apparent magnitude of $m\simeq24$ if it is
strongly-lensed, independent of its redshift.  This is brighter than
discussed above for GW170814, but still sufficiently faint to be
beyond the reach of most of the current generation of wide-field
searches for EM counterparts (Section~\ref{sec:strategy}).

\subsection{Low mass X-ray binaries}\label{sec:accretion}

We also consider the possibility of detecting an EM counterpart that
is much fainter than a kilonova.  A BH-BH merger in vacuum is not
expected to emit any EM radiation; however, numerous theoretical ideas
for EM counterparts to BH-BH mergers not in vacuum have been proposed
following the detection of GW150914 \citep[e.g.,][]{Li2016, Loeb2016,
  Lyutikov2016, Murase2016, Morsony2016, Perna2016, Woosley2016,
  Yamazaki2016, Bartos2017, Dai2017BBHEM, deMink2017, Janiuk2017,
  Ryan2017, Stone2017}.  By way of illustration, we adopt a simple
model of an EM counterpart, apply Equation~\eqref{eqn:frozen}, and
compare with the sensitivity of our search.

We speculate that low mass BH X-ray Binaries (LMXRB) during outburst
in the Milky Way could provide an illustrative upper limit on the
brightness of the EM counterparts to BH-BH merger.  The brightest
LMXRB seen to date is V404\,Cyg, with a BH mass of $9\Msol$ and peak
extinction corrected absolute $V$-band magnitude during outburst of
$M_V\simeq-4.7$ \citep[e.g.,][]{vanParadijs1994, Bernadini2016}.  We
further assume that the luminosity of the accretion disc is
proportional to the mass of the BH, and the accretion rate as a
fraction of the Eddington limit.  Therefore, combining these
assumptions with the dependence of BH mass on redshift (discussed in
Section~\ref{sec:intro}), gives the following expression for the
estimated absolute magnitude $M$ of the EM counterpart to a lensed
BH-BH merger:
\begin{equation}
  M=M_0-2.5\,\log\left(\frac{\Lambda}{\Lambda_0}\,.\,\frac{\mathbb{M}_{{\rm f},\mu=1}}{\mathbb{M}_0}\,.\,\frac{1+z_{\mu=1}}{1+z}\right),
  \label{eqn:cygx1}
\end{equation}
where $z$ is the true redshift of the lensed BH-BH, $\mathbb{M}_{{\rm
    f},\mu=1}$ is the final BH mass of a BH-BH merger inferred
assuming $\mu=1$, $\Lambda$ denotes accretion rate as a fraction of
the Eddington limit, and we adopt $M_0=-4.7$, $\mathbb{M}_0=9\Msol$
and $\Lambda_0=1$ for V404\,Cyg.  Substituting
Equation~\eqref{eqn:cygx1} in to Equation~\eqref{eqn:frozen}, and
adopting $\mathbb{M}_{\rm f,\mu=1}\simeq53\,\Msol$,
$z_{\mu=1}\simeq0.12$ for GW170814 \citep{GWTC-1}, a nominal source
redshift of $z\simeq1$, and $\Lambda=1$, gives $m\simeq33$.  This is a
factor $\simeq10^3-10^4$ fainter than the transient point sources that
we are able to recover in our difference imaging
(Section~\ref{sec:analysis}), which is unsurprising given that we did
not set out to detect such faint EM counterparts.  Moreover, to
underline how challenging any possible detection of an EM counterpart
to a BH-BH would be, $\Lambda\gs10$ would be required to bring the
apparent magnitude of a source within reach of an observation of depth
comparable with the Hubble Ultra Deep Field, i.e.\ $m\simeq30$.
Alternatively, follow-up observations of a GW170814-like source
initially placed at $D_{\rm L,\mu=1}=170\,{\rm Mpc}$ and with
$\Lambda=1$ would also be detectable at $m\simeq30$, based on this
speculative V404\,Cyg-like scenario.  Finally, in all of this
discussion, we have assumed that $k$-corrections are negligible, as
the absolute magnitudes are in the (effectively) rest-frame $V$-band
of sources at $z\simeq1$ that we observe in the $i$-band.


\section{Summary}\label{sec:conclusion}

In the nights immediately following the announcement of the detection
of GW170814, we observed two strong-lensing cluster cores --
Abell~3084 and SMACS\,J0304.3$-$4401 -- identified using the sky
localization available from the LVC.  Our observations were conducted
with the GMOS and MUSE instruments on the Gemini-South and Very Large
Telescope, respectively.  The data reach a sensitivity to point
sources of $m(5\sigma)\simeq\operatorname{25--26}$, and our search for
transient sources is sensitive down to $m=25$ in the continuum and
line fluxes of $5\times10^{-17}\,{\rm erg\,s^{-1}\,cm^{-2}}$.  We
detect no credible candidate transient sources in the data down to
these limits.  This is the most sensitive search to date for EM
counterparts to GW sources, independent of considerations of possible
lens amplification.

The lens magnification suffered by a lensed GW source cancels out the
inverse square law, and therefore the apparent magnitude of a point
source EM counterpart of given luminosity is set by the luminosity
distance inferred from the GW data assuming no lens magnification.
The apparent magnitude of EM counterparts to lensed GW sources is
independent of the true redshift of the source.  We therefore show, as
a proof of concept, that we can exclude the idea that GW170814 is a
NS-BH or NS-NS source at $z>8$ that is lensed by either of these
clusters.  We also show that observations with 30-m class and/or
space-based telescopes will be required to conduct meaningful searches
for lensed EM counterparts to BH-BH sources.  We will consider the
details of such observing strategies, and those required for lensed
NS-NS and NS-BH sources in a future article.

In summary, we have confirmed the feasibility of searching for EM
counterparts to candidate lensed GW sources with ground-based $8$-m
class telescopes, and described some important considerations for
future development of this new observing strategy within the rapidly
growing field of GW astronomy.

\section*{Acknowledgments}

\sloppypar GPS dedicates this article to the memory of Robert Smith,
and thanks Olivia Mueller for her support during August 2017.  GPS
also thanks Gary Mamon and colleagues at the Institut d'Astrophysique
de Paris for their warm welcome during the latter stages of writing
this article, and Alberto Vecchio for his advice and support.  We
thank Iair Arcavi and Leo Singer for assistance with
Figure~\ref{fig:skyloc}, Alyssa Drake for providing us with the
results of completeness tests to blindly detect line emitters within
MUSE datacubes using {\sc muselet}, and Chris Done and Sylvain Chaty
for helpful discussions about low mass X-ray binaries.  We thank the
staff and Directors of the Gemini and La Silla Paranal Observatories
for their superb support of our observing programmes and awards of
Director's Discretionary time respectively.  We acknowledge support
from the Science and Technology Facilities Council through the
following grants: ST/N000633/1 (GPS, MB, CPLB, WMF); ST/L00075X/1,
ST/P000451/1 (MJ); and ST/K005014/1 (JV).  JR acknowledges support
from the ERC starting grant 336736-CALENDS.  RJM acknowledges support
from the Royal Society.

\bibliographystyle{mn2e}


\label{lastpage}

\end{document}